\documentclass[english]{article}
\usepackage[utf8]{inputenc}
\usepackage{amstext}
\usepackage{amssymb}
\usepackage{graphicx}
\usepackage{esint}

\newcommand{\ug}{ \; = \; }

\newcommand{\infi}{\infty}

\newcommand{\bb}{\begin{equation}}
\newcommand{\ee}{\end{equation}}
\newcommand{\bega}{\begin{eqnarray}}
\newcommand{\ega}{\end{eqnarray}}
\newcommand{\begae}{\begin{eqnarray*}}
\newcommand{\egae}{\end{eqnarray*}}

\newcommand{\h}{\hspace*{4ex}}
\newcommand{\dis}{\displaystyle}

\newcommand{\om}{\omega}

\newcommand{\kr}{k_{\rho}}

\makeatletter

\newcommand{\address}[1]{
\par {\raggedright #1
\vspace{1.4em}
\noindent\par}
}

\usepackage{cite}

\makeatother

\usepackage{babel}

\begin{document}

\baselineskip 0.55cm

\title{Modeling of Space-Time Focusing of Localized Nondiffracting Pulses}


\author{Michel Zamboni-Rached$^{1,2}$ and  Ioannis M.
Besieris$^{3}$}


\maketitle

\address{$^{1}$University of Campinas, Campinas, SP, Brazil.

$^{2}$Department of Electrical and Computer Engineering at the
University of Toronto, Toronto, ON, Canada

$^{3}$The Bradley Department of Electrical and Computer
Engineering, Virginia Polytechnic Institute and State University,
Blacksburg VA 24060, USA}

\begin{abstract}
In this paper we develop a method capable of modeling the
space-time focusing of nondiffracting pulses. The new pulses can
possess arbitrary peak velocities and, in addition to being
resistant to diffraction, can have their peak intensities and
focusing positions chosen \emph{a priori}. More specifically, we
can choose multiple locations (spatial ranges) of space/time
focalization; also, the pulse intensities can be chosen in
advance. The pulsed wave solutions presented here can have very
interesting applications in many different fields, such as
free-space optical communications, remote sensing, medical
apparatus, etc.

\end{abstract}

\section{Introduction}

\h It is well known that the scalar wave equation, in particular,
and more generally Maxwell's equations have very interesting
classes of solutions named Nondiffracting Waves
\cite{shepp,britt,kiselev,sez,durnin,besi1,lu,donn,livro1,livro2},
also called Localized Waves. These beams and pulses are immune to
diffraction effects but have the drawback of containing infinite
energy. This problem, however, can be solved
\cite{sez,besi1,besi2,mrh,mrh2} and the finite-energy versions of
the ideal nondiffracting waves (INWs) can resist the diffraction
effects for long (finite) distances when compared to the ordinary
waves.


\h In this paper, we take a step forward in the theory of the
nondiffracting pulses by introducing a new method that enables us
to perform a space/time modelling on them. In other words, we can
construct new nondiffracting localized pulse solutions in such a
way that we can choose where and how intense their peaks will be
within a longitudinal spatial range $0 \leq z \leq L$. This is not
just an ordinary focusing method \cite{focx,focx1,focx2}, where
there is just one point of focalization; actually, it is a much
more powerful method because it allows the choice of multiple
locations (spatial ranges) of space/time focalization where the
pulse intensities also can be chosen \emph{a priori}.



\h Such modeling of the space/time evolution of ideal localized
nondiffracting pulses with subluminal, luminal or superluminal
peak velocities is made through a suitable and discrete
superposition of ideal standard nondiffracting pulses. The new
resulting waves can possess potential applications in many
different fields, such as free-space optical communications,
remote sensing, medical apparatus, etc.


\h In Section 2 we present important results related to ideal
nondiffracting pulses with azimuthal symmetry, with the
bidirectional and unidirectional decomposition approaches playing
important roles. Section 3 is devoted to the modeling of the
space-time focusing of localized nondiffracting pulses and
examples are presented to confirm the efficiency of the method.
Section 4 is devoted to the conclusions.


%
%

\section{Important points related to the ideal (standard) nondiffracting
pulses}

\h In this section we present several important points regarding
the ideal nondiffracting pulses, stressing the importance of the
bidirectional and unidirectional decomposition methods used to
deriving closed analytical solutions describing such waves. Here,
we do not enter into mathematical details and demonstrations, as
this subject has already been very well developed in a series of
papers \cite{besi1,besi2,mrh,mrh2}.

\h Considering only propagating waves and azimuthal symmetry, the
general solution to the homogeneous scalar wave equation, $ \Box
\psi = 0$ (with $\Box$ being the d'Alembertian), can be written in
cylindrical coordinates as:


\bb \psi(\rho,z,t) \ug
\int_{-\infi}^{\infty}d\om\int_{-\infty}^{\infty}dk_z\int_{0}^{\infty}\kr\,d\kr
\, A'(\kr,k_z,\om)\,\delta\left(\kr^2 - \left(\frac{\om^2}{c^2} -
k_z^2 \right)\right) J_0(\kr \rho)e^{ik_z z}e^{-i\om t}
\label{Sgeral1} \ee

with $A'(\kr,k_z,\om)$ being an arbitrary function and $\delta(.)$
the Dirac delta function. The integral solution (\ref{Sgeral1}) is
nothing but a superposition of zero-order Bessel beams involving
the angular frequency $(\om)$ and the transverse ($\kr$) and
longitudinal ($k_z$) wave numbers.

\h It is well known \cite{besi1,besi2,mrh,mrh2} that an ideal
nondiffracting pulse with peak velocity $0 \leq V \leq \infty$ can
be obtained when the spectrum $A'(\kr,k_z,\om)$ forces a coupling
of the type $\om = Vk_z + b$, with $b$ a constant, between the
angular frequency and the longitudinal wave number. Such a
coupling, and the solutions resulting from it, can be much more
easily obtained when the bidirectional or unidirectional
decompositions \cite{besi1,besi2,mrh,mrh2} are adopted. These are
characterized by changes from the spatial ($z$) and time ($t$)
coordinates to the new ones, $\zeta$ and $\eta$, given by

\bb \left\{\begin{array}{clr}
\zeta = z - Vt \\
\\
\eta = z + ut
\end{array} \right.  \label{decomp}
 \ee

with $V>0$ and $\forall u\neq -V$. The transformation
(\ref{decomp}) is called bidirectional decomposition when $u > 0$,
and unidirectional decomposition when $u \leq 0$.

\h In the new coordinates, the integral solution (\ref{Sgeral1}),
after an integration on $\kr$, can be written as:

\bb \psi(\rho,\zeta,\eta)  \ug
\int_{\alpha_{min}}^{\alpha_{max}}d\alpha\int_{\beta_{min}}^{\beta_{max}}d\beta
S(\alpha,\beta) J_0(\rho\,s(\alpha,\beta))\, e^{-i\beta
\eta}e^{i\alpha\zeta} \,\, , \label{solint} \ee

where

\bb s(\alpha,\beta) \ug  \dis{\sqrt{\left(\frac{V^2}{c^2} -1
\right)\alpha^2 + \left(\frac{u^2}{c^2} -1 \right)\beta^2 +
2\left(\frac{u V}{c^2} + 1 \right)\alpha\beta}} \ee

and $S(\alpha,\beta)$ is the spectral function, with the new
spectral parameters $\alpha$ and $\beta$ given by

\bb \left\{\begin{array}{clr}
\alpha = \dis{\frac{1}{u+V}}\,(\om + u k_z) \\
\\
\beta = \dis{\frac{1}{u+V}}\,(\om - V k_z)
\end{array} \right. \,\, , \label{decomp2}
 \ee

\h The limits of the integrals in Eq.(\ref{solint}) depend on the
values of $V$ and $u$ and define the allowed values of $\alpha$
and $\beta$ in order to avoid evanescent waves. In the plane
($\om,k_z$), these values correspond to the region between the
straight lines $\om = c k_z$ and $\om = - c k_z$. To make easier
the derivation of exact analytical solutions, we can choose a
subdomain of the allowed values, considering, for instance, the
first and fourth quadrants of the plane $(\om,k_z)$.




\h Now, with the integral representation (\ref{Sgeral1}) it is
quite simple to consider spectra that fix a relation of the type
$\om = Vk_z + b$ which, as we have said, yield ideal
nondiffracting pulses. For this, with the new spectral parameters,
we just need a new spectral function of the type

\bb S(\alpha,\beta) \ug \delta(\beta - \beta_0)A(\alpha,\beta)
\,\, , \label{specLW} \ee

with $\beta_0$ a constant. Due to the Dirac delta function
$\delta(\beta - \beta_0)$ in the spectrum, the integral expression
(\ref{solint}) becomes a superposition of zero-order Bessel beams
with $\om$ and $k_z$ lying on the straight line\footnote{This can
be understood by noting that the Dirac delta function forces the
condition $\beta = \beta_0 \rightarrow \om = Vk_z + (u+V)\beta_0$}
$\om = V k_z + (u+V)\beta_0$. The pulse peak velocity will be
given by $V$ and can be subluminal ($V<c$), luminal ($V=c$) or
superluminal ($V>c$). The value of $u$ in the decomposition
(\ref{decomp}) can be chosen to facilitate the analytical and
exact integration of (\ref{solint}).

\begin{figure}[!h]
\begin{center}
\scalebox{.4}{\includegraphics{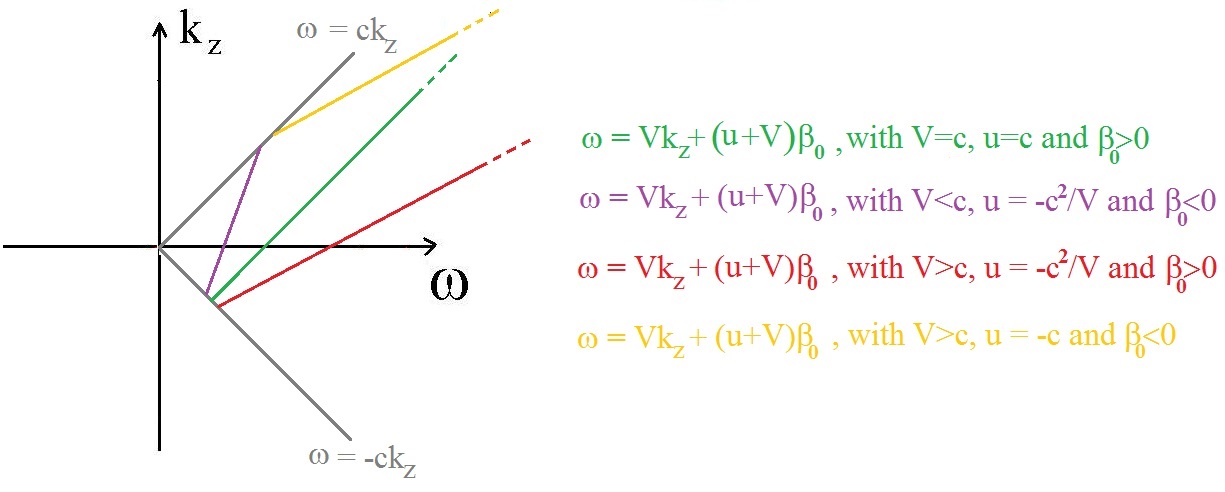}}
\end{center}
\caption{The semi-straight lines and the line segment given by
$\beta=\beta_0 \rightarrow \om = Vk_z + (u+V)\beta_0$ which define
the different types of ideal nondiffracting pulses. The values
chosen for $u$ are those that make easier the exact integration of
Eq.(\ref{solint}).} \label{retas}
\end{figure}

\h On the first and fourth quadrants of the plane $(\om,k_z)$,
Fig.(\ref{retas}) shows the semi-straight lines and the
line-segment given by $\beta=\beta_0$ which define the different
types of ideal nondiffracting pulses.

\h By using Eq.(\ref{specLW}), with $\beta_{min} < \beta_0 <
\beta_{max}$, in (\ref{solint}) we get

\bb \psi(\rho,\zeta,\eta)  \ug e^{-i\beta_0 \eta}
\int_{\alpha_{min}}^{\alpha_{max}}d\alpha A(\alpha,\beta_0)
J_0(\rho\,s(\alpha,\beta_0))\, e^{i\alpha\zeta} \,\, .
\label{solint2} \ee

In general, the spectra $A(\alpha,\beta_0)$ used in the Localized
Wave theory yield pulses centered on $\zeta = 0 \rightarrow z=Vt$.
Below we list several important nondiffracting pulses obtained
from Eq.(\ref{solint2}).

\begin{itemize}
    \item {Subluminal MacKinnon-type pulse:

    \bb \psi(\rho,\zeta,\eta) \ug \mathcal{N}\,e^{-i \beta_0 \eta}\,
     {\rm sinc}\left(\frac{c}{V}|\beta_0|\sqrt{\gamma^{-2}\rho^2 -
     (a_1+i\zeta)^2\,}\right)\,\, , \label{Mack}\ee
%

where $a_1>0$, $\gamma = (1-V^2/c^2)^{-1/2}$ and $\mathcal{N} = -2
a_1 c \beta_0 \exp(a_1 c \beta_0/V)/[V(1 - \exp(2 a_1 c
\beta_0/V))]$.

    This pulse \cite{besi2,mack,salo1,saari1,sub} is obtained from Eq.(\ref{solint2}) considering
    $V<c$, $u = -c^2/V$, $\beta_0 < 0$, $\alpha_{min} = c\beta_0/V$, $\alpha_{max} = -c\beta_0/V$
    and $A(\alpha,\beta_0) = -V\mathcal{N}/(2 c \beta_0)\exp(a_1\alpha)$.


    It is possible to show \cite{sub} that in order to minimize the contribution of
    the backward components to this solution it is necessary that
    $a_1>>V/[|\beta_0(c+V)|]$. It is also possible to show that, with this condition on $a_1$,
    the central angular frequency of this pulse is $\om_c \approx c^2(V/c+1)|\beta_0|/V$,
    with a frequency bandwidth $\Delta\om \approx V/a_1$.

    The modulus square of the Mackinnon pulse is undistorted
    during the propagation, with a spot size of radius $\Delta\rho_0 =
    \gamma\sqrt{(\pi V / |\beta_0| c)^2 + a_1^2}$.}

    \item {The Luminal Focus Wave Mode (FWM):

  \bb \psi(\rho,\zeta,\eta) \ug a_1 \frac{e^{-i \beta_0\eta}}{a_1 - i\zeta}
  \exp\left(-\frac{\beta_0\rho^2}{a_1 - i\zeta}\right) \label{fwm}    \ee

    This pulse \cite{besi1} is obtained from Eq.(\ref{solint2}) considering
    $V=c$, $u = c$, $\beta_0 > 0$, $\alpha_{min} = 0$, $\alpha_{max} =
    \infty$ and $A(\alpha,\beta_0)= a_1 \exp(-a_1\alpha)$, with $a_1>0$ a constant.

    It is possible to show \cite{tese1,tese2} that the central frequency of this pulse is $\om_c = c\beta_0 + c/2a_1$, with a bandwidth $\Delta\om
    \approx c/a_1$. It is also possible to show \cite{besi1,tese1,tese2} that in order to minimize the contribution of
    the backward (wave) components of this solution (i.e., to make it causal) it is necessary that
    $a_1 << 1/\beta_0$, which implies that $\Delta\om \approx
    2\om_c$. Such wideband frequency spectrum is a necessary condition to
    ensure the causality of the FWM pulse. This
    may give the wrong idea that luminal nondiffracting pulses must be
    ultrashort. Actually, although it is the case for the FWM solution, the
    existence of completely causal luminal nondiffracting pulses with arbitrary frequency bandwidths
    is quite possible and the only issue is that we do not know exact
    analytical solutions for these cases, which of course can be
    obtained through numerical simulations.

    The modulus square of the FWM pulse is undistorted
    during the propagation, with a spot size of radius $\Delta\rho_0 = \sqrt{a_1/(2\beta_0)} $.}

    \item {The Superluminal Focus Wave Mode (SFWM):

  \bb \psi(\rho,\zeta,\eta) \ug \mathcal{N}\frac{e^{-i\beta_0\eta}}{\sqrt{\gamma'^{-2}\rho^2 + (a_1-i\zeta)^2}}
  \exp{\left(-\frac{c}{V}\beta_0\sqrt{\gamma'^{-2}\rho^2 +
  (a_1-i\zeta)^2}\right)}\,\, , \label{xfwm} \ee

  where $a_1>0$, $\gamma' = (V^2/c^2 -1)^{-1/2}$ and $\mathcal{N} = a_1\exp(a_1 c \beta_0/V )$. This pulse \cite{besi2,mrh,mrh2} is obtained from Eq.(\ref{solint2}) considering
    $V>c$, $u = -c^2/V$, $\beta_0 > 0$, $\alpha_{min} = c |\beta_0|/V$, $\alpha_{max}   =
    \infty$ and $A(\alpha,\beta_0)= \mathcal{N}\exp(-a_1\alpha)$.

    It is possible to show \cite{besi1,besi2,mrh} that to minimize the contribution of
    the backward components to this solution it is necessary that
    $a_1 << V/[(V-c)\beta_0]$. It is also possible to show that the central frequency
    of this pulse is $\om_c = (1 - c/V)c\beta_0 + V/(2a_1)$, with an angular frequency
    bandwidth $\Delta\om =  V/a_1 $ which, due to the condition on $a_1$, can be approximated by
    $\Delta\om \approx 2 \om_c$. The ultra wideband frequency spectrum also occurs here but,
    as in the luminal case, it is not a fundamental characteristic of superluminal
    nondiffracting pulses as we will see in the next case.

    The modulus square of the SFWM pulse is undistorted
    during the propagation, with a spot size of radius $\Delta\rho_0 =
    \gamma'\sqrt{[a_1 + V/(2 c \beta_0)]^2 - a_1^2}$.}

    \item {The Superluminal Focus Wave Mode totally free
of backward components:

  \bb \psi(\rho,\zeta,\eta) \ug a_1 X e^{-i\beta_0\eta}\exp{\left[-\frac{\beta_0}{\frac{V}{c}+1}
  \left(a_1 -i\zeta - X^{-1}\right)\right]}\,\, , \label{mrh2}  \ee

  where $X = [\gamma'^{-2}\rho^2 + (a_1 - i\zeta)^2]^{-1/2}$ is the
  classical X-wave solution, $a_1>0$ and $\gamma' = (V^2/c^2
  -1)^{-1/2}$. This pulse \cite{mrh2} is obtained from Eq.(\ref{solint2}) considering
    $V>c$, $u = -c$, $\beta_0 < 0$, $\alpha_{min} = 0$, $\alpha_{max} =
    \infty$ and $A(\alpha,\beta_0)= a_1 \exp(-a_1\alpha)$. It possesses a central frequency
    $\om_c = c|\beta_0| + V/(2a_1)$, with an angular frequency bandwidth $\Delta\om = V/a_1$. It is
important to note that this pulse solution is totally free of
backward components which is a great advantage. Due to this, it
can possess any time width, not necessarily characterized by
wide-frequency bandwidths as it is usually the case with other
nondiffracting pulses of known analytical solutions.

    The modulus square of this pulse is undistorted
    during the propagation, with a spot size of radius $\Delta\rho_0 \approx
    \gamma'a_1\sqrt{(V/c+1)^2/(4\beta_0^2a_1^2) + (V/c+1)/(a_1|\beta_0|)}$
    in the case $a_1|\beta_0|>(V/c+1)/(2\sqrt{2}-2)$, otherwise we have $\Delta\rho_0 \approx
    \gamma'a_1$.}

\end{itemize}

\h The intensity of all the pulses described above are undistorted
for all time. Actually, this is the basic characteristic of all
known ideal nondiffracting pulses (and beams) which, as we have
already said, possess infinite energy. The latter drawback can be
addressed by truncating \cite{ziol1,lwtruncada,mrb} the ideal
solutions\footnote{In this case the resulting field is given by
diffraction integrals, which rarely can be solved analytically}
(finite aperture generation) or by concentrating the spectrum
$A'(\kr,k_z,\om)$ entering Eq.(\ref{Sgeral1}) in a region
surrounding the straight line $\om = Vk_z + b$ instead of
collapsing it exactly over that line \cite{besi1,besi2,mrh,mrh2}.
In both cases, the resulting waves attain finite energy and are
resistant to the diffraction effects for long (but not infinite)
distances.

\h In the next section we are going to take a step forward in the
theory of the Localized Waves by introducing a new method that
will enable us to model the space-time focusing of ideal
nondiffracting pulses.

\

\section{The method for modelling the space-time focusing of ideal nondiffracting pulses}

\h In general, the peak of an ideal nondiffracting pulse with
azimuthal symmetry, $\psi(\rho,\zeta,\eta)$, occurs at

\bb \rho=0 \;\; {\rm and}\;\; \zeta = 0 \;\; ({\rm i.e.,}\;
z=Vt)\; \rightarrow \eta = \left(1+\frac{u}{V} \right)z \equiv
\left(1+\frac{u}{V} \right)z_p \,\, , \ee

where we have used Eq.(\ref{decomp}) and called $z=z_p$, $z_p$
being the peak's z-position. In this case, the pulse's peak based
on Eq.(\ref{solint2}) is described by

\bb \psi(\rho=0,\zeta=0,\eta=(1+u/V)z_p)  \ug
\exp\left[-i\left(1+\frac{u}{V}\right)\beta_0 z_p\right]
\int_{\alpha_{min}}^{\alpha_{max}}d\alpha A(\alpha,\beta_0)
\label{solint3} \ee

\h Next, a new pulse solution, $\Psi$, will be constructed in such
a manner that both the position of its peak, as well as the
intensity of the peak, can be chosen within a longitudinal spatial
range $0 \leq z \leq L$. The new pulse is defined as a
superposition of LW pulses like those in Eq.(\ref{solint2}), with
the same spectral function $A(\alpha,\beta_0)$ but with values of
$\beta$ different from $\beta_0$, in such a way that we can write
for the new solution:

\bb \Psi(\rho=0,\zeta=0,\eta=(1+u/V)z_p)  \equiv \sum_{n=-N}^{N}
B_n \, U_n \exp\left[-i\left(1+\frac{u}{V}\right)\beta_n
z_p\right] \label{solint4} \ee

where $U_n = \int_{\alpha_{min}}^{\alpha_{max}}d\alpha
A(\alpha,\beta_n)$, with the coefficients $B_n$ and the values of
$\beta_n$ yet unknown.

\h If the choices

\bb \beta_n = \frac{1}{1+u/V}\left(Q + \frac{2\pi}{L}n\right)
\label{betan} \ee

and

\bb  B_n = \frac{1}{L\,U_n} \int_{0}^{L} F(z_p) e^{i\frac{2\pi
n}{L}z_p} d z_p \,\, ,  \label{Bn} \ee

are made, with $Q$ and $F(z_p)$ a constant and an arbitrary
function, respectively, the intensity of the expression in
Eq.(\ref{solint4}) within $0 \leq z_p \leq L$ will result in:

\bb |\Psi(\rho=0,\zeta=0,\eta=(1+u/V)z_p)|^2 = |F(z_p)|^2 \ee

and so we can chose, through the function $F(z_p)$, the locations
and how intense the pulse's peak will appear.

\h The 3D \emph{exact solution} is given by:

\bb \Psi(\rho,\zeta,\eta) \ug   \sum_{n=-N}^{N} B_n
\psi_n(\rho,\zeta,\eta)\,\, , \label{Psi} \ee

where

\bb \psi_n(\rho,\zeta,\eta) \ug e^{-i\beta_n \eta}
\int_{\alpha_{min}}^{\alpha_{max}} A(\alpha,\beta_n)
J_0(\rho\,s(\alpha,\beta_n))\, e^{i\alpha\zeta} \label{psin}  \ee

\h This approach is different from the ordinary space/time
focusing methods, as those developed in \cite{focx,focx1,focx2},
where there is just one point of focalization. It is a much more
powerful method because, we repeat, it allows us to choose
multiple locations (spatial ranges) of space/time focalization,
where the pulse intensities also can be chosen \emph{a priori}.
This technique can be seen as a pulsed version of the Frozen Wave
method \cite{fw1,fw2,fw3,fw4,fw5,fw6} (originally developed for
beams) and it can be implemented for subluminal, luminal and
superluminal pulsed solutions. It is important to note that within
all focal lines, the resulting pulses will be nondiffracting.



\h In general, the nondiffracting pulses $\psi_n$ given in
Eq.(\ref{psin}), and used in the new solution (\ref{Psi}), have
all their $\beta_n$ with the same signal, positive or negative.
Once we have chosen the values of $\beta_0$ [which in turn defines
the value of $Q$ via Eq.(\ref{betan})] and $L$, this fact will
imply a maximum value allowed for $N$, which restricts the number
$2N+1$ of terms in the fundamental superposition (\ref{Psi}). The
maximum value of $N$ can be obtained from Eq.(\ref{betan}), with
the requirement that the signal of all $\beta_n$ has to be equal
to the signal for the chosen $\beta_0$.

%
%
%

\h It is not difficult to show that in any case the maximum value
of $N$ in (\ref{Psi}) has to obey:

\bb N \leq \frac{|1+u/V|}{2\pi}\, L \, |\beta_0|   \ee


\h With the method in hand, a natural question emerges: Which
criteria should be used in choosing the values of the parameters
$\beta_0$, which determines $Q$ through Eq.(\ref{betan}), $L$ and
any other appearing within the spectral function
$A(\alpha,\beta_n)$ in Eq.(\ref{psin}), which defines $\psi_n$?

\h First we have to say that any new solution
$\Psi(\rho,\zeta,\eta)$, given by Eq.(\ref{Psi}), will have
several important characteristics associated to it, such as a
central frequency, frequency bandwidth and the transverse spot
size, which will be approximately similar to those presented by
the central pulse, $\psi_0$, of the superposition (\ref{Psi}).
These characteristics of $\psi_0$ are described by equations
involving the parameters $\beta_0$ and those occurring in
$A(\alpha,\beta_0)$, which, therefore, can be determined once the
desired characteristics mentioned above are chosen\footnote{In
general, just the central frequency and the frequency bandwidth or
just the central frequency and the spot-size are necessary to
determine the values of $\beta_0$ and other possible parameters
occurring in $A(\alpha,\beta_0)$}.


\h The value of $L$ cannot be evaluated in this way, but a good
criterion is to choose it taking as reference the diffraction
length, $Z_{gauss}$ of a Gaussian pulse possessing the same spot
size of the desired new pulse. Good choices for $L$ would be those
limited to values not much greater than hundred or thousand times
the value of $Z_{gauss}$, otherwise we risk dealing with
unrealistic situations in experimental apparatus sizes to generate
the finite energy version of the pulse.

\h Having presented our method, we are going to illustrate it with
a few examples, which will confirm its efficiency and simplicity.

\h It is important to notice that in the following examples some
of the desired pulse's peak evolution patterns involve step
functions, which are discontinuous. However, the resulting pulses
will not present any discontinuity because they are given by a
discrete and finite superposition of continuous pulsed solutions
[see Eq.(\ref{Psi})]. Actually, what we are going to get are
resulting pulses that approach the desired patterns, maintaining
however the necessary properties of continuity and
differentiability.

\emph{\textbf{First example:}}

\h Here, we shall model the space-time focusing of a subluminal
nondiffracting pulse. To do this, we are going to use the
fundamental solution (\ref{Psi}), with the $\psi_n$ given by the
subluminal MacKinnon solution in Eq.(\ref{Mack}), with $\beta_0$
replaced with $\beta_n$.

\h In this case, we can set the parameters $\beta_0$ [which
defines $Q$ through Eq.(\ref{betan}] and $a_1$ according to the
central frequency and frequency bandwidth of the resulting pulse,
remembering that these characteristics are approximately similar
to those of the central pulse $\psi_0$ of the superposition
(\ref{Psi}). In this case, $\psi_0$ is given by the Mackinnon-type
solution, Eq.(\ref{Mack}), whose characteristics cited above were
presented in the previous section.

\h Considering a peak velocity $V= 0.999c$, with $u=-c^2/V$, a
central angular frequency $\om_c = 2.98\times10^{15}$rad/s and a
frequency bandwidth $\Delta\om \approx 10^{-2}\om_c$, we can get
$\beta_0 = -4.96 \times 10^{6}{\rm m}^{-1}$ and $a_1 = 1.01 \times
10^{-5}$m. With this, the intensity spot radius for the resulting
pulse can be estimated as being $\Delta\rho_0 \approx 0.226$mm.
Concerning the value of $L$, which defines the range $ 0 \leq z_p
\leq L $ within which the space-time modelling will be made, we
note that an ordinary (Gaussian) pulse with the same spot size
considered here would possess a diffraction length
$Z_{diff}=0.14$m so, according to our previous considerations, the
value of $L$ should not be many orders of magnitude greater than
this value for avoiding unrealistic situations. Let us choose
$L=30Z_{diff} \approx 4.2$m.

\h For the function $F(z_p)$, whose modulus square will shape the
positions and intensity of the resulting pulse's peak in the range
$0 \leq z_p \leq L$, let us choose a ladder intensity pattern on
the space interval $L/3 \leq z_p \leq 2L/3$, and zero outside it.
More specifically, we wish that within $0 \leq z \leq L$ the
pulse's peak intensity obeys
$|\Psi(\rho=0,\zeta=0,\eta=(1+u/V)z_p)|^2 = |F(z_p)|^2$, with


\bb
 F(z_p) \ug \left\{\begin{array}{clr}
 1 \;\;\; & {\rm for}\;\;\; l_1 < z_p < l_2  \\
 \\
 \sqrt{2} \;\;\; & {\rm for}\;\;\; l_2 < z_p < l_3  \\
\\
 \sqrt{3}\;\;\; & {\rm for}\;\;\; l_3 < z_p < l_4 \\
\\
 0 \;\;\; & {\rm otherwise}\;\;\; \,\, ,
\end{array} \right. \label{Fz1}
 \ee

Here, $l_1=L/3$, $l_2=L/3 + \Delta l$, $l_3=L/3 + 2\Delta l$ and
$l_4=L/3 + 3\Delta l$, with $\Delta l = L/9$.

\h The resulting pulse is given by Eq.(\ref{Psi}) with the
coefficients $B_n$ given by Eq.(\ref{Bn}) and $\psi_n$ given by
Eq.(\ref{Mack}) through the replacement $\beta_0 \rightarrow
\beta_n$, with $\beta_n$ given by Eq.(\ref{betan}). Here, the
maximum allowed value for $N$ is 6,616, but we will use $N=60$.

\begin{figure}[!h]
\begin{center}
\scalebox{.55}{\includegraphics{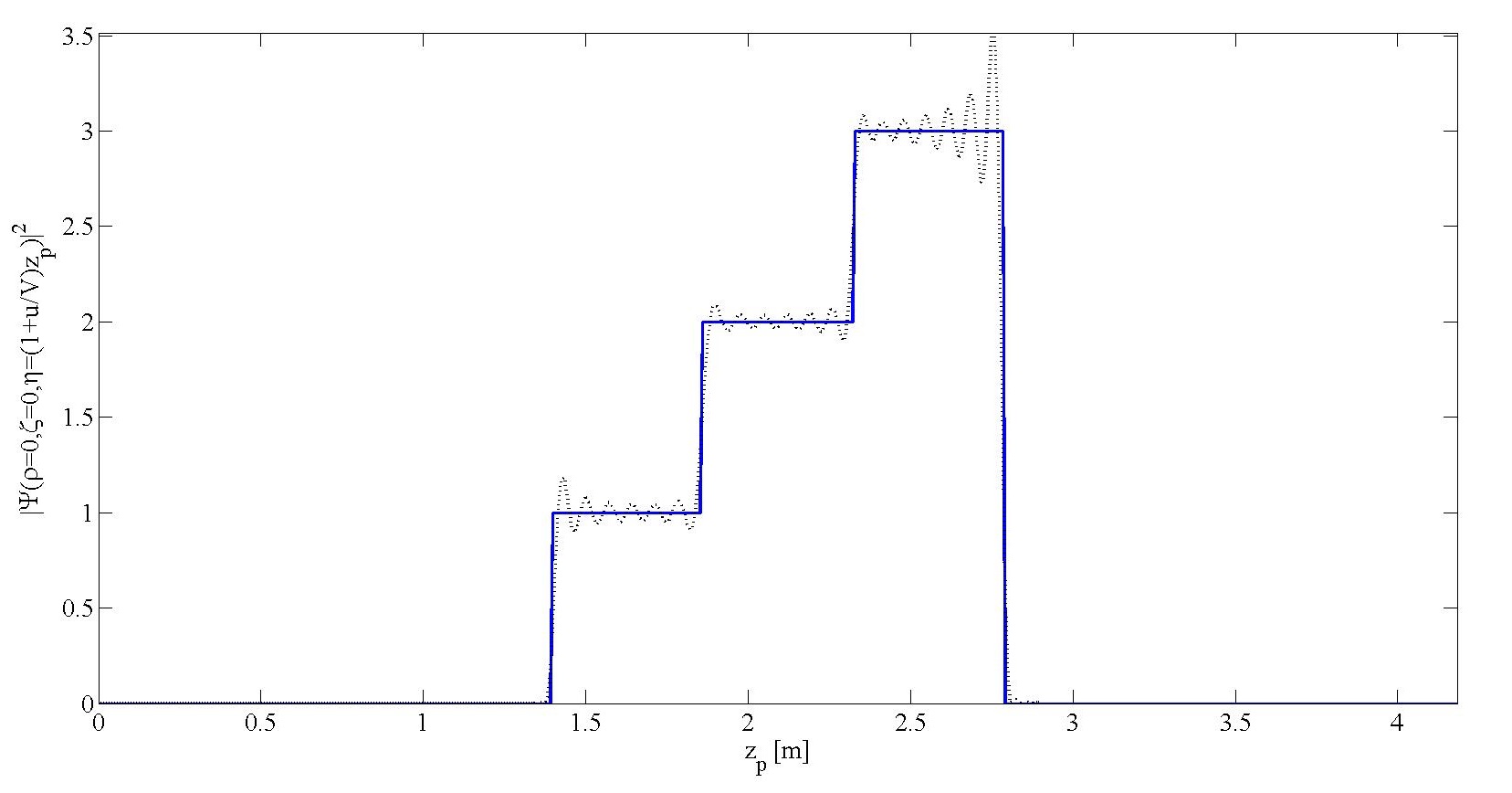}}
\end{center}
\caption{The peak intensity evolution,
$|\Psi(\rho=0,\zeta=0,\eta=(1+u/V)z_p)|^2$, of the resulting
subluminal pulse is shown in doted line, while the desired peak
intensity evolution for it, $|F(z_p)|^2$, is shown in continuous
line. A good agreement between them is observed.} \label{Fig2}
\end{figure}

\h Figure \ref{Fig2} shows, within the range $0 \leq z_p \leq L$,
the evolution of the actual pulse's peak intensity,
$|\Psi(\rho=0,\zeta=0,\eta=(1+u/V)z_p)|^2$, in doted line and the
desired spatial evolution for it, $|F(z_p)|^2$, in continuous
line. We can see a good agreement between them.

\begin{figure}[!h]
\begin{center}
\scalebox{.65}{\includegraphics{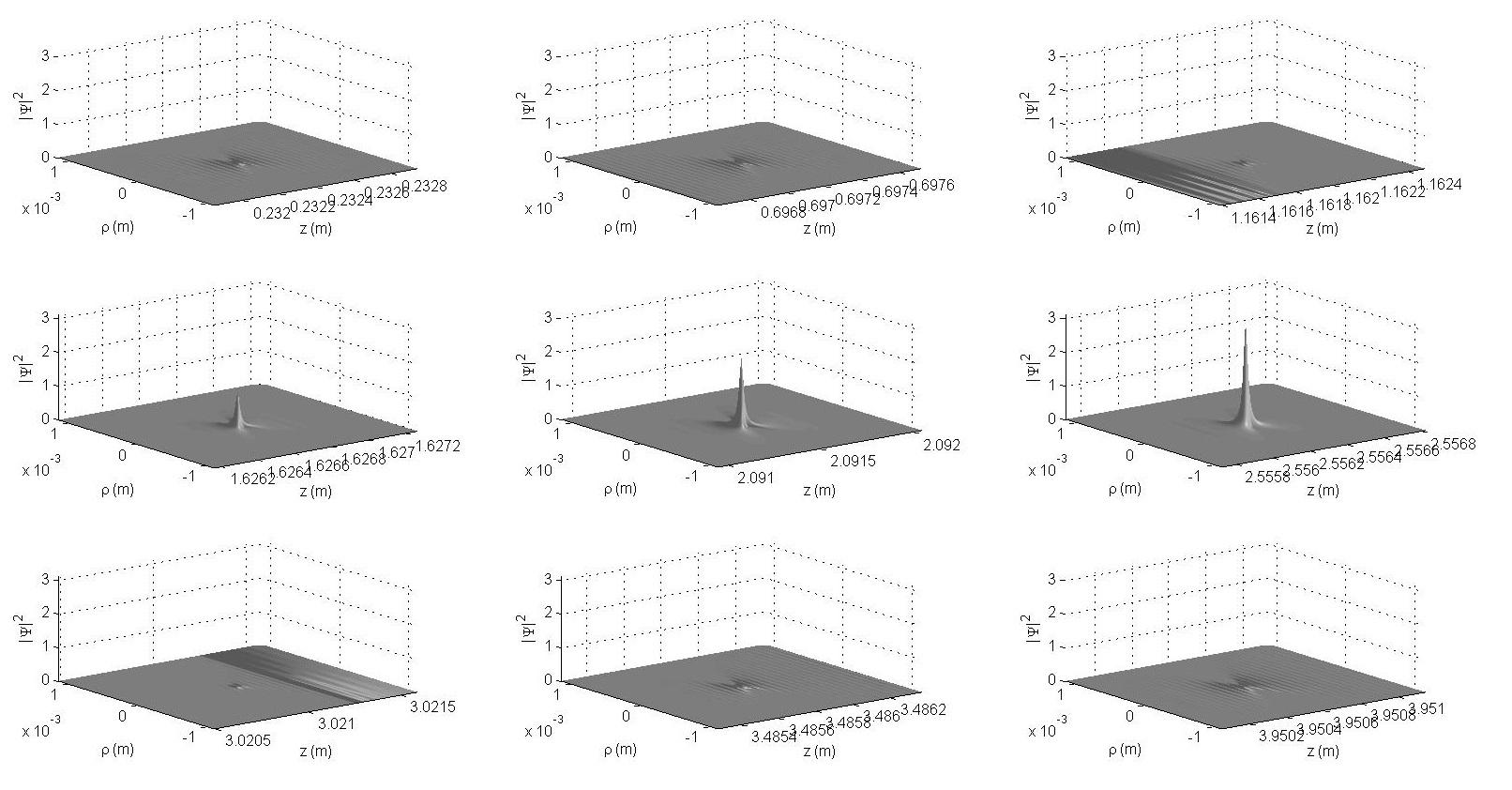}}
\end{center}
\caption{The $3D$ pulse intensity, $|\Psi(\rho,\zeta,\eta)|^2$,
for the resulting subluminal pulse at nine different instants of
time. The first, second and third lines of the subfigures show the
pulse evolution within the ranges $0<z<L/3$, $L/3<z<2L/3$ and
$2L/3<z<L$, respectively. We can see that the resulting
nondiffracting pulse possesses the desired space-time focusing
characteristics.} \label{Fig3}
\end{figure}

\h Even more interesting is Fig.(\ref{Fig3}), which shows the $3D$
pulse intensity -- $|\Psi(\rho,\zeta,\eta)|^2$ -- at nine
different instants of time. More specifically, the first, second
and third lines of the subfigures show the pulse evolution within
the ranges $0<z<L/3$, $L/3<z<2L/3$ and $2L/3<z<L$, respectively.
It is very clear that the resulting nondiffracting pulse possesses
the desired space-time focusing characteristics.

\newpage

\emph{\textbf{Second example:}}

\h Next, we are going to model the space-time focusing of a
luminal nondiffracting pulse. For this, we will use our
fundamental solution (\ref{Psi}) with the $\psi_n$ given by the
luminal FWM solution, Eq.(\ref{fwm}), with $\beta_0$ replaced by
$\beta_n$. .


\h In this case, the parameters $\beta_0$ and $a_1$ are chosen
according to the desired central frequency and transverse
spot-size for the resulting pulse, these characteristics being
approximately similar to those of the central pulse $\psi_0$ of
the superposition (\ref{Psi}). In this case, $\psi_0$ is given by
the FWM solution, Eq.(\ref{fwm}), of which the characteristics
cited above were presented in the previous section.

\h Considering for the resulting pulse a peak velocity $V = c$,
with $u=c$, a central angular frequency $\om_c =
2.98\times10^{15}$rad/s and an intensity spot radius $\Delta\rho_0
= 10\mu$m, we obtain $\beta_0 = 2.518 \times 10^{2} {\rm m}^{-1}$
and $a_1 = 5.036 \times 10^{-8}$m. In this case the frequency
bandwidth is $\Delta\om \approx 2\om_c$. Furthermore, we choose
$L=0.516$m, which is $300$ times greater than the diffraction
length of a Gaussian pulse with the same spot size considered
here.


\h Within the region $0 \leq z_p \leq L$ the pulse's peak
intensity evolution obeys
$|\Psi(\rho=0,\zeta=0,\eta=(1+u/V)z_p)|^2 = |F(z_p)|^2$, with the
choice

\bb
 F(z_p) \ug \left\{\begin{array}{clr}
 1/\sqrt{2} \;\;\; & {\rm for}\;\;\; 0 < z_p < L/3  \\
 \\
 0 \;\;\; & {\rm for}\;\;\; L/3 < z_p < 2L/3  \\
\\
 1 \;\;\; & {\rm for}\;\;\; 2L/3 < z_p < L \,\, ,
\end{array} \right. \label{Fz2}
 \ee

The resulting pulse is given by Eq.(\ref{Psi}) with the
coefficients $B_n$ given by Eq.(\ref{Bn}) and $\psi_n$ given by
Eq.(\ref{fwm}) through the replacement $\beta_0 \rightarrow
\beta_n$, with $\beta_n$ given by Eq.(\ref{betan}). Here, we use
$N=30$.

\begin{figure}[!h]
\begin{center}
\scalebox{.55}{\includegraphics{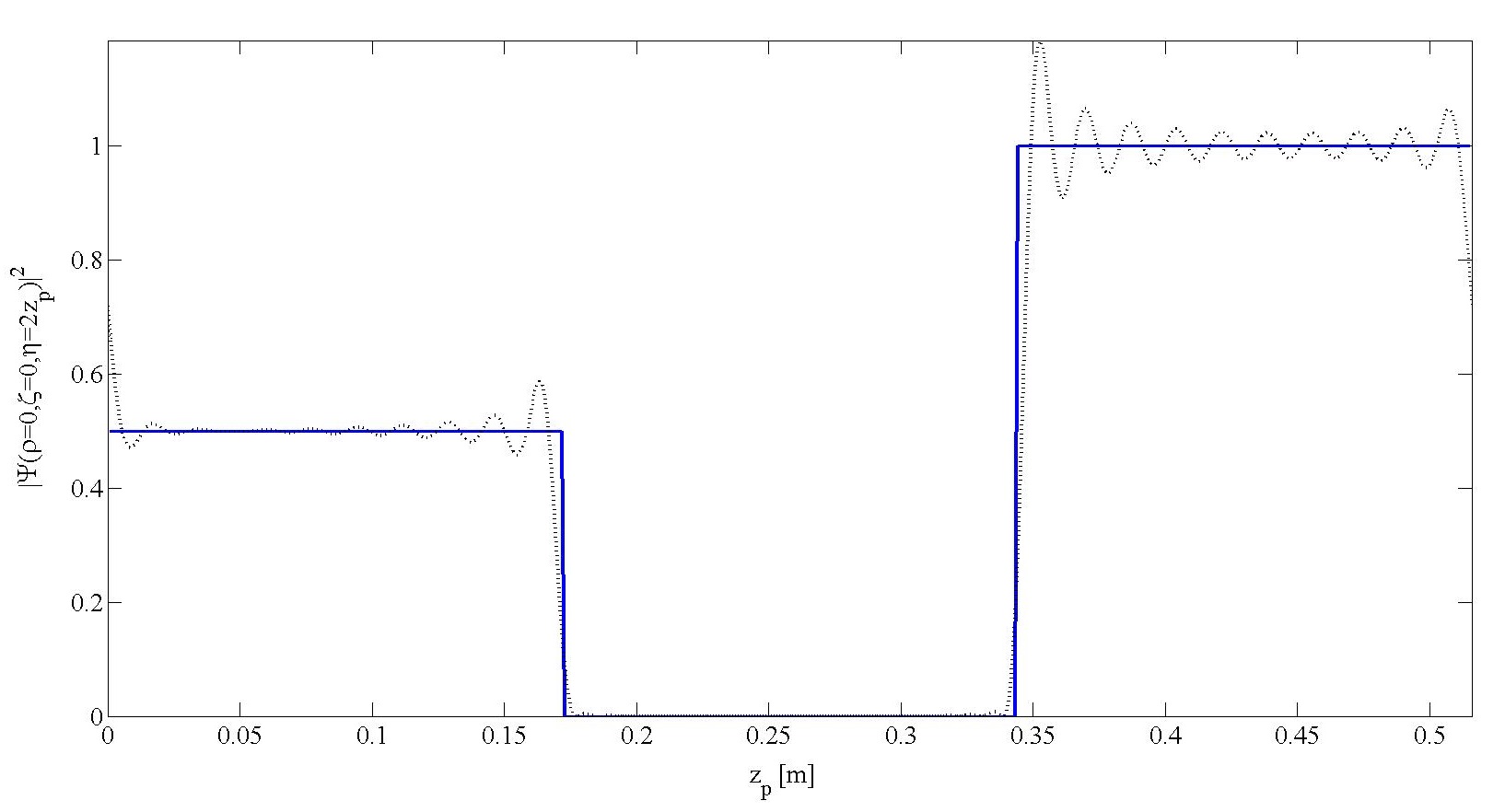}}
\end{center}
\caption{The peak intensity evolution,
$|\Psi(\rho=0,\zeta=0,\eta=(1+u/V)z_p)|^2$, of the resulting
luminal pulse is shown in doted line, while the desired peak
intensity evolution for it, $|F(z_p)|^2$, is shown in continuous
line. A good agreement between them is seen.} \label{Fig4}
\end{figure}

\h Figure \ref{Fig4} shows, within the range $0 \leq z_p \leq L$,
a good agreement between the actual pulse's peak intensity
evolution -- $|\Psi(\rho=0,\zeta=0,\eta=(1+u/V)z_p)|^2$ -- (doted
line) and the desired spatial evolution -- $|F(z_p)|^2$ -- for it
(continuous line).

\begin{figure}[!h]
\begin{center}
\scalebox{.65}{\includegraphics{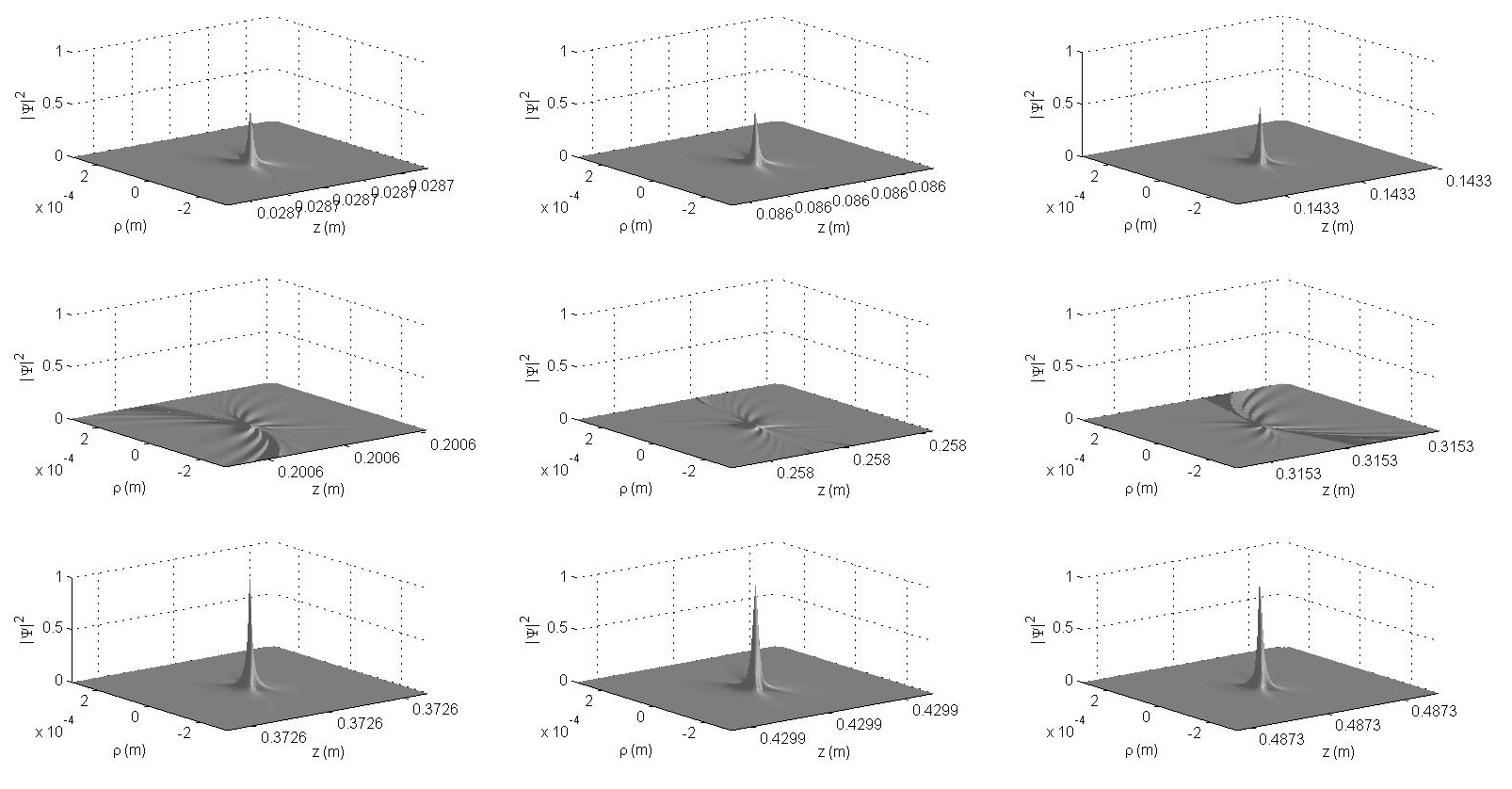}}
\end{center}
\caption{The $3D$ pulse intensity, $|\Psi(\rho,\zeta,\eta)|^2$,
for the resulting luminal pulse at nine different instants of
time. The first, second and third lines of the subfigures show the
pulse evolution within the ranges $0<z<L/3$, $L/3<z<2L/3$ and
$2L/3<z<L$, respectively. We can see that the resulting
nondiffracting pulse possesses the desired space-time focusing
characteristics.} \label{Fig5}
\end{figure}

\h Figure (\ref{Fig5}) shows the $3D$ evolution of the pulse
intensity -- $|\Psi(\rho,\zeta,\eta)|^2$ -- at nine different
instants of time. The first, second and third lines of subfigures
show the pulse evolution within the ranges $0<z<L/3$, $L/3<z<2L/3$
and $2L/3<z<L$, respectively. It is clear that the resulting
nondiffracting pulse possesses the desired space-time focusing
characteristics.

\newpage

\emph{\textbf{Third example:}}

\h Finally, we shall model the space-time focusing of a
superluminal nondiffracting pulse. Again, we use our fundamental
solution (\ref{Psi}), but now with the $\psi_n$ given by the
superluminal solution in Eq.(\ref{mrh2}) with $\beta_0$ replaced
with $\beta_n$.


\h Considering a peak velocity $V = 1.0001 c$, with $u=-c$, a
central angular frequency $\om_c = 2.98\times10^{15}$rad/s, a
frequency bandwidth $\Delta\om = \om_c/10^3$, and following the
same procedure as in the previous examples, but now considering
$\psi_0$ as given by Eq.(\ref{mrh2}), we can get $\beta_0 =
-9.9242\times 10^{6}{\rm m}^{-1}$ and $a_1 = 1.0072\times
10^{-4}$m. In this case, the intensity spot radius for the
resulting picosecond pulse can be estimated as being $\Delta\rho_0
\approx 0.32$mm. We also choose $L=111.2$m, which is approximately
$400$ times greater than the diffraction length of a Gaussian
pulse with the same spot size considered here.

\h For the space-time focusing modelling within $0 \leq z \leq L$,
we wish the superluminal pulse's peak intensity
$|\Psi(\rho=0,\zeta=0,\eta=(1+u/V)z_p)|^2 = |F(z_p)|^2$ to behave
as

\begin{figure}[!h]
\begin{center}
\scalebox{.55}{\includegraphics{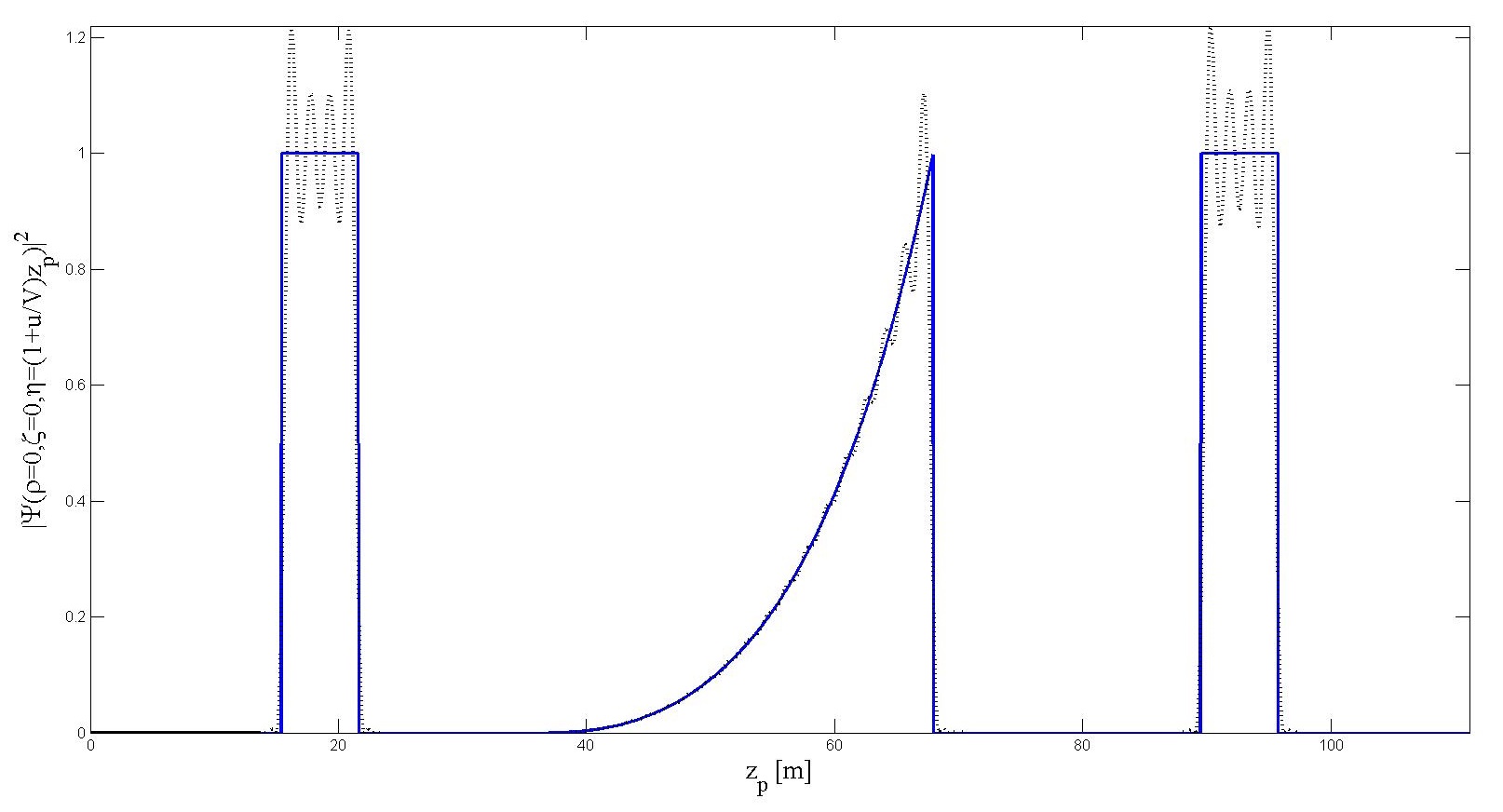}}
\end{center}
\caption{The peak intensity evolution,
$|\Psi(\rho=0,\zeta=0,\eta=(1+u/V)z_p)|^2$, of the resulting
superluminal pulse is shown in doted line, while the desired peak
intensity profile for it, $|F(z_p)|^2$, is shown in continuous
line. A good agreement between them is observed.} \label{Fig6}
\end{figure}

\bb
 F(z_p) \ug \left\{\begin{array}{clr}
 1 \;\;\; & {\rm for}\;\;\; l_1 < z_p < l_2  \\
 \\
 \dis{\frac{\exp[(z-l_3)/(l_4-l_3)]-1}{e-1}}
 \;\;\; & {\rm for}\;\;\; l_3 < z_p < l_4  \\
\\
 1 \;\;\; & {\rm for}\;\;\; l_5 < z_p < l_6  \,\, ,
\end{array} \right. \label{Fz3}
 \ee

where $l_1=1.25 \Delta l$, $l_2=1.75 \Delta l$, $l_3=3 \Delta l$,
$l_4= 5.5 \Delta l$, $l_5=7.25 \Delta l$ and $l_6=7.75 \Delta l$,
being $\Delta l = L/9$. This desired pattern for the evolution of
the pulse's peak intensity is shown by Fig.(\ref{Fig6}) in
continuous line.

\h The resulting pulse is given by Eq.(\ref{Psi}) with the
coefficients $B_n$ given by Eq.(\ref{Bn}) and $\psi_n$ given by
Eq.(\ref{mrh2}) through the replacement $\beta_0 \rightarrow
\beta_n$, with $\beta_n$ given by Eq.(\ref{betan}). Here, we use
$N=70$.


\h Figure \ref{Fig6} shows, in doted line, the actual pulse's peak
intensity profile along the axis,
$|\Psi(\rho=0,\zeta=0,\eta=(1+u/V)z_p)|^2$, which presents a good
agreement with the desired pattern (continuous line).

\begin{figure}[!h]
\begin{center}
\scalebox{.65}{\includegraphics{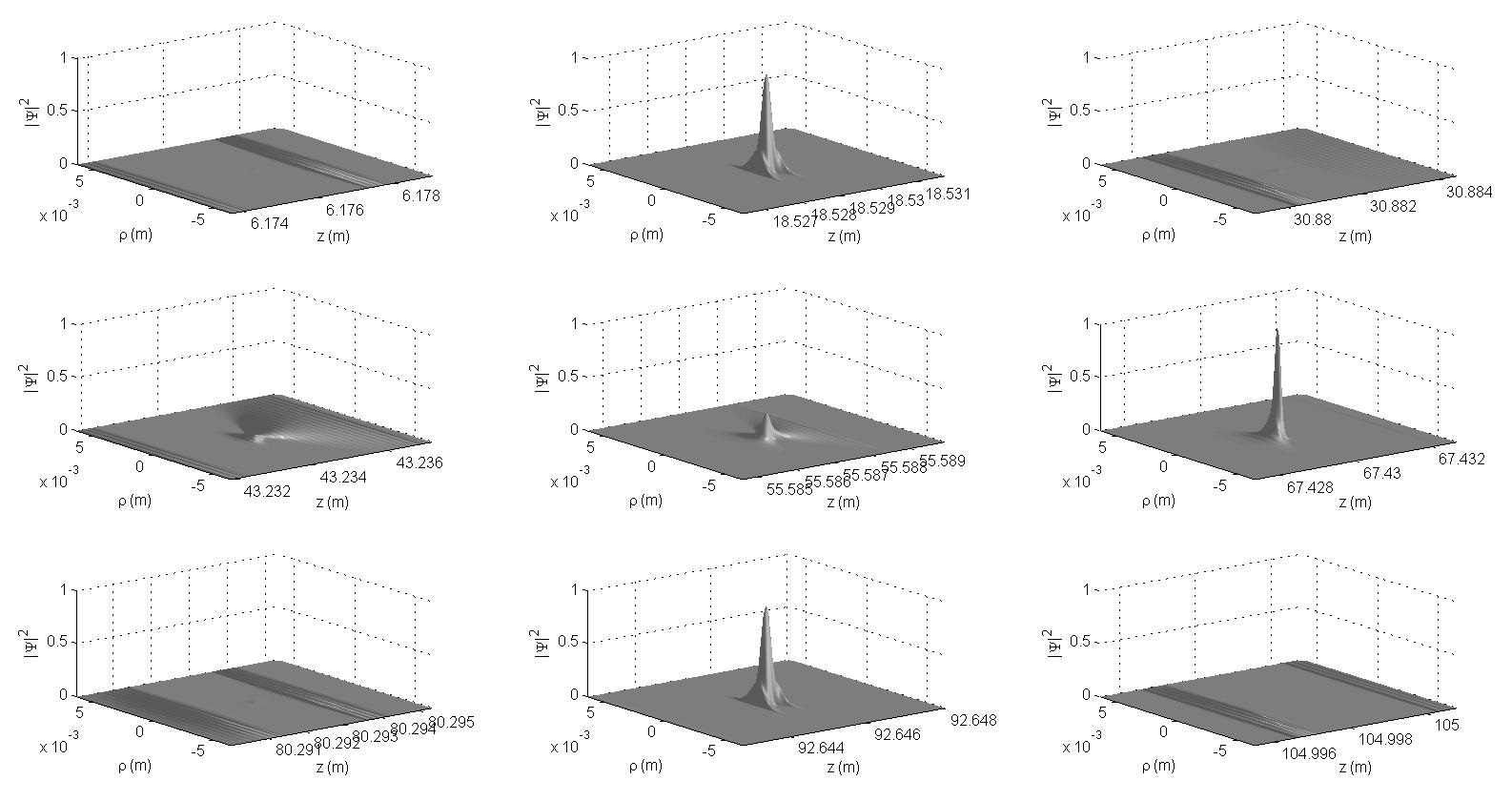}}
\end{center}
\caption{The $3D$ pulse intensity, $|\Psi(\rho,\zeta,\eta)|^2$,
for the resulting superluminal pulse at nine different instants of
time. The first, second and third lines of the subfigures show the
pulse evolution within the ranges $0<z<L/3$, $L/3<z<2L/3$ and
$2L/3<z<L$, respectively. We can see that the resulting
nondiffracting pulse possesses the desired space-time focusing
characteristics.} \label{Fig7}
\end{figure}

\h Figure (\ref{Fig7}) shows the $3D$ evolution of the pulse
intensity -- $|\Psi(\rho,\zeta,\eta)|^2$ -- at nine different
instants of time. The first, second and third lines of subfigures
show the pulse evolution within the ranges $0<z<L/3$, $L/3<z<2L/3$
and $2L/3<z<L$, respectively. Again, it is clear that the
resulting nondiffracting pulse possesses the desired space-time
focusing characteristics.

\newpage

\section{Conclusions}

\h In this paper, a novel method has been presented that enables
the modeling of space/time focusing of nondiffracting pulses.
Specifically, it has been shown how to construct new localized
pulse solutions in such a way that one can choose where and how
intense their peaks will be within a longitudinal spatial range $0
\leq z \leq L$. This approach can be considered a step forward in
the Localized Wave theory as it enables one to take advantage of
the great potential of the nondiffracting wave pulses in a highly
selective manner, by choosing \emph{a priori} multiple locations
(spatial ranges) of space/time focalization, with the pulse
intensities also being chosen.

\h Space/time focusing modeling of ideal localized nondiffracting
pulses with subluminal, luminal or superluminal peak velocities
has been carried out by means of suitable superpositions of ideal
standard nondiffracting pulses. The resulting new waves can have
potential applications in many different fields, such as
free-space optical communications, remote sensing, medical
apparatus, etc.

\h The pulse solutions resulting from the space/time focusing
method possess infinite energy content as they are obtained from
discrete superpositions of ideal nondiffracting pulses. The finite
energy version of our method will be presented elsewhere.

\section*{Acknowledgements}

\h The authors thank Mo Mojahedi for valuable discussions and kind
collaboration.

\h This work was supported by FAPESP (under grant 2013/26437-6);
CNPq (under grants 312376/2013-8).

%
%
%

\end{document}